\documentclass[12pt,letter,doublespace]{article}
\usepackage{amssymb}
\usepackage{natbib,graphicx,setspace,lscape,longtable}
\usepackage{natbib,epsfig,graphicx}
\usepackage{amsmath,amsthm,amssymb}
\usepackage{color,xcolor}
\usepackage{threeparttable}
\usepackage{changes}

\usepackage{booktabs}

\usepackage{hyperref}

\RequirePackage[mathlines, displaymath]{lineno}
\usepackage{multirow}
\usepackage{graphicx}
\usepackage{subfig}
\usepackage{threeparttable}
\DeclareGraphicsExtensions{.eps, .jpg, .pdf}
\bibpunct{(}{)}{;}{a}{,}{,}

\def\beqr{\begin{eqnarray}}
\def\eeqr{\end{eqnarray}}
\def\beqrs{\begin{eqnarray*}}
\def\eeqrs{\end{eqnarray*}}
\def\bep{\begin{prop}}
\def\eep{\end{prop}}
\def\bc{\begin{center}}
\def\ec{\end{center}}

\newcommand{\bG}{{\bf G}}
\newcommand{\bt}{\mathbf{t}}
\newcommand{\bs}{\mathbf{s}}
\newcommand{\bu}{\mathbf{u}}

\newcommand{\bH}{\mathbf{H}}

\newcommand{\bP}{\mathbf{P}}

\newcommand{\trans}{^{\mbox{\tiny{T}}}}

\numberwithin{equation}{section}

\def \bec{\begin{center}}
\def \enc {\end{center}}
\def \bee {\begin{eqnarray*}}
\def \ene {\end{eqnarray*}}

\def \bear{\begin{array}}
\def \enar{\end{array}}

\def \bs{\begin{slide}}
\def \es{\end{slide}}

\newcommand{\ggamma}{\boldsymbol{\gamma}}
\newcommand{\ttheta}{\boldsymbol{\theta}}

\newcommand{\SSigma}{\boldsymbol{\Sigma}}

\newcommand{\ddelta}{\boldsymbol{\delta}}

\newcommand{\xxi}{\boldsymbol{\xi}}
\newcommand{\XXi}{\boldsymbol{\Xi}}

\newcommand{\eeta}{\boldsymbol{\eta}}

\newcommand{\LLambda}{\boldsymbol{\Lambda}}

\newcommand{\UUpsilon}{\mbox{\boldmath $\Upsilon$}}

\newcommand{\bA}{{\bf A}}

\newcommand{\bD}{{\bf D}}

\newcommand{\bU}{{\bf U}}

\newcommand{\bY}{{\bf Y}}
\newcommand{\bZ}{{\bf Z}}
\newcommand{\bW}{{\bf W}}
\newcommand{\bM}{{\bf M}}
\newcommand{\bS}{{\bf S}}
\newcommand{\bQ}{{\bf Q}}

\numberwithin{equation}{section}  

\newtheorem{thm}{Theorem}[section]

\newtheorem{prop}{Proposition}[section]

\usepackage{algorithm}
\usepackage{algorithmicx}

\renewcommand{\baselinestretch}{1.23}
\begin{document}

\title{ Multi-Kink Quantile Regression for Longitudinal Data with Application to the Progesterone Data Analysis}

\author{{ Chuang Wan$^{1}$, Wei Zhong$^{1,*}$, Wenyang Zhang$^{3}$ and Changliang Zou$^{4}$}\\
{$^1$ Xiamen University; $^2$The University of York; $^3$ Nankai University}\\
$^*$Email: {wzhong@xmu.edu.cn}\\
First version: December 20, 2020}

\maketitle{}

\begin{abstract}
Motivated by investigating the relationship between progesterone and the days in a menstrual cycle in a longitudinal study, we propose a multi-kink quantile regression model for longitudinal data analysis.  It relaxes the linearity condition and assumes different regression forms in different regions of the domain of the threshold covariate. In this paper, we first propose a multi-kink quantile regression for longitudinal data. Two estimation procedures are proposed to estimate the regression coefficients and the kink points locations: one is a computationally efficient profile estimator under the working independence framework while the other one considers the within-subject correlations by using the unbiased generalized estimation equation approach. The selection consistency of the number of kink points and the asymptotic normality of two proposed estimators are established.  Secondly, we construct a rank score test based on partial subgradients for the existence of kink effect in longitudinal studies.  Both the null distribution and the local alternative distribution of the test statistic have been derived.  Simulation studies show that the proposed methods have excellent finite sample performance.  In the application to the longitudinal progesterone data, we identify two kink points in the progesterone curves over different quantiles and observe that the progesterone level remains stable before the day of ovulation, then increases quickly in five to six days after ovulation and then changes to stable again or even drops slightly.
\end{abstract}


\pagestyle{plain}
\setcounter{page}{1}

\section{Introduction}

Longitudinal data are frequently observed in many fields such as clinical medicine, biomedical science, social sciences and economics.  In longitudinal data, the within-subject correlations of repeated measurements bring many challenges to both parameter estimation and statistical inference.  In traditional regression analysis such as linear regression, the impacts of covariates on a response are often assumed to be constant on the whole domain of the covariates.  However, this assumption may be invalid in some applications. \cite{li2015quantile} developed the bent line quantile regression \citep{li2011bent} for the longitudinal framework and showed that the cognitive decline was gradual like normal aging in the preclinical stage of Alzheimer's disease (AD) and then distinguishably accelerated as the disease progresses after a certain change point.  The bent line regression \citep{li2011bent} or the kink regression \citep{hansen2017regression} assumes that  different regression forms are separately modeled on two sides of an unknown threshold but still continuous at the threshold.  \cite{li2011semiparametric} also found that there exists a threshold effect between the blood pressure change and the progression of microalbuminuria among individuals with type-I diabetes using censored longitudinal data. \cite{ge2020threshold} proposed a threshold linear mixed model to determine the cutpoint of a continuous regressor and to estimate the interaction effect between the treatment and subgroup indicator on longitudinal responses.

All these aforementioned methods assume there is only one single change point.  This assumption could not be satisfied in applications.  Without prior knowledge on the single kink point assumption, it is more reasonable to assume a general regression model with multiple kink points for the longitudinal data. \cite{das2016fast} introduced a likelihood-based estimation approach for a broken-stick model with multiple change points for longitudinal data. However, the number of change points is assumed to be known as a priori. \cite{zhong2021estimation} considered a multi-kink quantile regression for the independent data.  However, the model does not directly apply to longitudinal data.
As \cite{li2015quantile} mentioned, ``\emph{Extensions to handle two or more change-points in the model are prohibited by algorithmic issues and merit future research.}"  This motivates us to study the multi-kink quantile regression (MKQR) for longitudinal data analysis.

In this paper, we propose a multi-kink quantile regression (MKQR) model for longitudinal data.  It assumes different regression forms in different regions of the domain of the threshold covariate.
The multi-kink regression can be considered as a special case of partial linear regression where the nonlinear relationship between the response and the threshold covariate is captured by a continuous piecewise linear model. Compared with nonparametric models, such as spline regression, the multi-kink design has the better interpretability by detecting the kink points locations and maintaining linear regressions in each segment of the threshold covariate.  We summarize main contributions as follows.  First, the MKQR model allows the covariate effects and kink points to vary across different quantiles and is robust to outliers and heavy-tailed errors simultaneously. Two estimation procedures are proposed to estimate the regression coefficients and the kink points. One is a computationally efficient two-step estimation procedure under the working independence framework where the within-subject correlations are ignored in the estimation step. The other one is a  generalized estimating equation (GEE) estimator \citep{liang1986longitudinal} to incorporate the correlation information within subjects. We estimate the number of kink points by transforming it into a model selection problem based on a quantile information criterion. The selection consistency of the number of kink points and the asymptotic normality of the estimators are established.
Second, we construct a rank score test based on partial subgradients for the existence of kink effect at a given quantile level for longitudinal data analysis.  Both the null distribution and the local alternative distribution of
the test statistic have been theoretically studied.  Third, we apply the proposed MKQR model to the longitudinal progesterone data to clearly identify two distinguishable kink points in the progesterone curves over different quantiles.  We observe that the progesterone level remains stable before the day of ovulation, then increases quickly in five to six days after ovulation and then changes to stable again or even drops slightly after the second kink point. Last, new R functions  for longitudinal data analysis are developed in the R package \emph{MultiKink} to implement the proposed estimation and inference procedures.

The rest of this paper is organized as follows.  In Section 2, we introduce the multi-kink quantile regression for the longitudinal data  and two parameter estimation procedures. The asymptotic properties are studied in Section 3.
Section 4 presents a quantile score-type test for the existence of kink points. Intensive simulation studies are conducted in Section 5 to evaluate the finite sample performances of the proposed methods. The longitudinal progesterone data analysis is included in Section 6.  A concluding remark is given in Section 7.  The technical proofs and additional simulation results are provided in the Appendix.

\section{Model and Methodologies}
\subsection{Model setting}
Let $Y_{ij}$ denote a response of interest, $X_{ij}$  be a univariate threshold variable and $\bZ_{ij}$ be a $p$-dimensional additional covariates  at the $j$th observation for the $i$th subject, where $j=1,2,\cdots,n_i$ and $i=1,2,\cdots,N$. Let $n=\sum_{i=1}^{N}n_i$ be the total number of observations. Denote $\bW_{ij}=(X_{ij},\bZ_{ij}\trans)\trans$.  Without loss of generality, we assume  $X_{ij}$ has a bounded support set $[M_1,M_2]$. At a given quantile level $\tau\in (0,1)$, define the $\tau$th condition quantile of $Y$ given $\bW$ as $Q_Y(\tau|\bW)=F^{-1}(\tau|\bW)=\inf\{u:F(u|\bW)\geq\tau\}$, where $F(\cdot|\bW)$ is the conditional density of $Y$ given $\bW$.

We consider a more flexible multi-kink quantile regression (MKQR) with an undetermined number of kink points for the longitudinal data,
\begin{equation}\label{eq:30}
Y_{ij}=\alpha_{0,\tau,0}+\alpha_{1,\tau,0}X_{ij}+\sum_{k=1}^{K_0}
\beta_{k,\tau,0}(X_{ij}-t_{k,\tau,0})I(X_{ij}>t_{k,\tau,0})+{\bf Z}_{ij}\trans{\ggamma}_{\tau,0}+e_{ij}^{(\tau)},
\end{equation}
where $M_1< t_{1,\tau}<t_{2,\tau}<\cdots<t_{K_0,\tau}<M_2$ are $K_0$ kink points and $\beta_{k,\tau,0}\neq0$ for $k=1,\cdots,K_0$ is the difference in slopes for the adjacent $k$th and $(k+1)$th regimes. There are totally $K_0+1$ regimes divided by $K_0$ kink points.
$e_{ij}^{(\tau)}$ is the error term with the $\tau$th quantile being zero conditional on $X_{ij}$ and $\bZ_{ij}$.
In the longitudinal models, $e_{ij}^{(\tau)}$'s are usually independent across subjects but correlated within a subject.
It is worth emphasizing that although the slope of $X_{ij}$ is discontinuous at $t_{\tau}$, but the regression function is continuous everywhere on the whole domain of $X$.

Denote $\eeta_0({\tau})=(\alpha_{0,\tau,0},\alpha_{1,\tau,0},
\beta_{1,\tau,0},\cdots,\beta_{K_0,\tau,0},\ggamma_{\tau,0}\trans)\trans$ as the vector of the true  regression coefficients and $\bt_0(\tau)=(t_{1,\tau,0},\cdots,t_{K_0,\tau,0})\trans$ as the vector of true kink points. Although all parameters including $K_0$ depends on the quantile level $\tau$, we will suppress such dependence on $\tau$ for ease of notations.  Model (\ref{eq:30}) can be re-expressed as the conditional quantile form
\begin{equation}\label{eq:4}
Q_Y(\tau;\ttheta_0|\bW_{ij})=\alpha_{0,0}+
\alpha_{1,0}X_{ij}+
\sum_{k=1}^{K_0}\beta_{k,0}
(X_{ij}-t_{k,0})_++\bZ_{ij}\trans\ggamma_0,
\end{equation}
where $\ttheta_0=(\eeta_0\trans,\bt_0\trans)\trans$ and  $(a)_+=aI(a>0)$ for any $a\in \mathbb{R}$. As pointed by \cite{li2015quantile}, model (\ref{eq:4}) means that 100$\tau$\% of the subjects have an outcome value no longer than $\alpha_{0,0}+
\alpha_{1,0}X_{ij}+
\sum_{k=1}^{K_0}\beta_{k,0}
(X_{ij}-t_{k,0})_++\bZ_{ij}\trans\ggamma_0$. 



\subsection{Parameter estimation}
We first introduce two estimation procedures to estimate the regression coefficients and the kink points given a fixed number of kink points. Then, we estimate the true number of kink points using a model selection approach based on a quantile information criterion.
\subsubsection{A working independence framework}
 To estimate $\ttheta=(\eeta\trans,\bt\trans)\trans$, a computationally efficient way is to  ignore possible correlations among measurements within subjects and to minimize the following objective function
\begin{equation}\label{eq:sn}
S_n(\ttheta)=n^{-1}\sum_{i=1}^N\sum_{j=1}^{n_i}\rho_\tau\{Y_{ij}-Q_Y(\tau;\ttheta|\bW_{ij})\},
\end{equation}
where $n=\sum_{i=1}^Nn_i$ is the total number of all observations and $\rho_\tau(v)=v\{\tau-I(v<0)\}$ is the quantile check function. It is  not easy to directly minimize  (\ref{eq:sn}) since the objective function  is neither differentiable nor convex. To this end, we propose a two-step profile estimation strategy to estimate both regression coefficients and kink location parameters simultaneously. The estimation steps are listed as below.
\begin{itemize}
  \item {\bf Step 1:}  Given a fixed number of kink points $K$, the dimension of $\ttheta$ is also determined. The profile estimator of $\eeta$ conditional on  $\bt$ is obtained by
\begin{equation}\label{eq:step1}
\widehat{\eeta}^I(\bt)=\underset{\eeta\in\mathcal{B}}{\arg\min}S_n\left(
\eeta,\bt\right),
\end{equation}
where $\mathcal{B}\subset\mathbb{R}^{2+p+K}$  is a compact set for $\eeta$.
  \item \noindent{\bf Step 2:} The estimator for $\bt$  at a given $K$ is therefore defined as
\begin{equation}\label{eq:step2}
\widehat{\bt}^I=\underset{\bt\in\LLambda}{\arg\min}S_n\left\{
\widehat{\eeta}^I(\bt),
\bt\right\}
\end{equation}
where $\LLambda=(M_1+\epsilon\leq t_1<t_2<\cdots<t_K\leq M_2-\epsilon)$ is a constrained  region for $\bt$ and $\epsilon$ is a small positive number to avoid edge effect.  Minimization of (\ref{eq:step2}) is a linearly constrained optimization which can be implemented by the adaptive barrier algorithm. 
\end{itemize}
\noindent
The two-step profile estimation procedure with ignoring the within-subject correlations is computationally efficient even when $K$ is relatively large. We denote the estimators under the working independence framework as $\widehat{\ttheta}^{I}=\left(\widehat{\eeta}^{I\mbox{\tiny{T}}}(\widehat{\bt}^I),
\widehat{\bt}^{I\mbox{\tiny{T}}}\right)\trans$.

\subsubsection{Incorporating correlations by using GEE}
Since the ignored within-subject correlations may incur some estimation efficiency loss, we then consider an efficient generalized estimating equation (GEE) approach \citep{liang1986longitudinal} to incorporate the within-subject correlations of longitudinal data. We estimate $\ttheta$ through the estimating equation
\begin{equation}
\sum_{i=1}^N\mathcal{X}_i(\ttheta)\bA_i^{-1}\psi_\tau
\{\bY_i-\bQ_{i}(\tau;\ttheta|\bW_{i})\}={\bf0},
\end{equation}
where $\bY_i=(Y_{i1},\cdots,Y_{in_i})\trans$, $\mathcal{X}_{i}(\ttheta)=\left(\mathcal{X}_{i1}(\ttheta),
\cdots, \mathcal{X}_{in_i}(\ttheta)\right)\trans$ with $\mathcal{X}_{ij}(\ttheta)=(1,X_{ij},(X_{ij}-t_1)_+,
\cdots,(X_{ij}-t_K)_+,\bZ_{ij}\trans,-\beta_1I(X_{ij}>t_1),\cdots,-\beta_KI(X_{ij}
>t_K))\trans$,   $\psi_\tau(u)=\tau-I(u<0)$, $\psi_\tau(\bu_i)=(\psi_\tau(u_{i1}),\cdots,\psi_\tau(u_{in_i}))\trans$  and  $\bQ_i(\tau;\ttheta|\bW_i)=(Q_{Y}(\tau;\ttheta|\bW_{i1}),\cdots,Q_{Y}(\tau;\ttheta|
\bW_{in_i}))\trans$.  $\bA_i$ is an $n_i\times n_i$ working correlation matrix to account for the correlations among observations within the $i$th subject. However, the efficiency of the GEE method  relies on the correct specification of $\bA_i$, and it will loss the estimation efficiency once $\bA_i$ is misspecified. To overcome this issue, we apply the quadratic inference functions (QIF) method \citep{qu2000improving} to characterize $\bA_i$ by a linear combination of basic matrices using
$$
\bA_i^{-1}=\sum_{l=1}^va_l\bM_{li},
$$
where $\bM_{li}$ is some given basic matrices, $a_l$'s are unknown constants, $l=1,\ldots,v$. As pointed by \cite{qu2000improving}, the basic matrices family should be rich enough to accommodate or  at least approach the true correlation structures. For example, if the working correlation has the AR(1) structure, then $\bA_i^{-1}$ can be represented by a combination of three basis matrices $\bM_{1i}$, $\bM_{2i}$ and $\bM_{3i}$, where $\bM_{1i}$ is the identity matrix, $\bM_{2i}$ has 1 on the two main subdiagonals and 0 elsewhere, and $\bM_{i3}$ has 1 on (1,1) and $(n_i,n_i)$ components and 0 elsewhere.

In the QIF approach, we do not need to  directly estimate the nuisance parameters $a_l$'s to explicitly specify the correlation structure. Instead, we consider multiple sets of estimating equations based on basic matrices to estimate $\ttheta$,  for $l=1,\cdots,v$,
 \begin{equation}\label{eq:basic}
 \sum_{i=1}^N \bS_{il}(\ttheta)
 =\sum_{i=1}^N \mathcal{X}_i(\ttheta)\trans\bM_{li}\psi_\tau\{\bY_i-
\bQ_i(\tau;\ttheta|\bW_i)\}
 =0.\end{equation}
Since the number of estimation equations is greater than the number of parameters, we apply the idea of the generalized method of moments (GMM)  \citep{hansen1982large} to estimate $\ttheta$ by combining multiple sets of estimating equations,
 \begin{equation}\label{eq:eta}
 \widehat{\ttheta}^{C}=\arg\min_{\ttheta}\bP_N(\ttheta)\equiv
 \arg\min_{\ttheta}\bS(\ttheta)\XXi_N^{-1}(\ttheta)\bS(\ttheta),
 \end{equation}
 where $\bS(\ttheta)=N^{-1}\sum_{i=1}^N \bS_i(\ttheta)$ with $\bS_i(\ttheta)=(\bS_{i1}\trans(\ttheta),\cdots,
\bS_{iv}\trans(\ttheta))\trans$, and $\XXi_N(\ttheta)$ is the covariance matrix of $\bS_i(\ttheta)$ which can be estimated by ${N}^{-1}\sum_{i=1}^N\bS_i(\ttheta)\bS_i\trans(\ttheta)-\bS(\ttheta)
 \bS\trans(\ttheta).$ From the computational aspect,  we apply the induced smoothing technique in \cite{leng2014smoothing} for (\ref{eq:eta}) and solve the smoothed objective function using the Newton-Raphson method.
 The detailed algorithm and the asymptotical equivalence are included in the Web Appendix A.

\subsection{Determine the number of kink points}
In real data analysis, the true number of kink points is usually unknown in practice and needs to be identified. In this subsection, we consider a model selection procedure to consistently estimate the true number of kink points.
With slight abuse of notation, we impose the  subscript ``$k$'' for parameter vectors to emphasize the MKQR model with $k$ kink points in the estimation  parts.
 By letting $k=0,1,\cdots,K^*$, we can obtain $K^*+1$ candidate models estimation results by using previous estimation procedures, where  $K^*$ is a pre-specified maximum number of kink points. Selecting an optimal $K$ can be transformed to  a model selection problem. We consider the  Schwarz-type quantile information criterion:
\begin{equation}\label{eq:sw}
\text{SIC}(k)=\log\left\{S_n(\widehat{\ttheta}_{k})\right\}+\frac{\log(n)}{2n}\omega_k,
\end{equation}
where $\widehat{\ttheta}_k$ is the estimates for $\ttheta_k$ using one of two estimation procedures, and  $\omega_k=2+p+2k$ is the length of parameter vectors.  Thus, the estimator for $K$ is $\widehat{K}=\underset{k=0,1,\cdots,K^*}{\arg\min}\text{SIC}(k)$.  Once  $K$ is determined, the final estimator for  $\ttheta$ is also obtained consequently.

\section{Asymptotic Properties}
\subsection{Asymptotic Normality}
We first study the limiting distribution of the profile estimator $\widehat{\ttheta}^{I}$ under the working independence framework.  Define the following two matrices:
\begin{eqnarray*}
\bH_n
=n^{-1}\sum_{i=1}^N\sum_{j=1}^{n_i}f_{ij}\{Q_Y(\tau;\ttheta_0|\bW_{ij})
\}\mathcal{X}_{ij}(\ttheta_0)
\mathcal{X}_{ij}(\ttheta_0)\trans,
\end{eqnarray*}
where $f_{ij}\{Q_Y(\tau;\ttheta_0|\bW_{ij})\}$ is the conditional density function of $Y_{ij}$ given $\bW_{ij}$; and
\begin{eqnarray}\nonumber
\SSigma_n(\ddelta)=n^{-1}\Bigg\{\sum_{i=1}^N\sum_{j=1}^{n_i}\tau(1-\tau)
\mathcal{X}_{ij}(\ttheta_0)\mathcal{X}_{ij}(\ttheta_0)\trans+
 \sum_{i=1}^N \sum_{j\neq j^{'}}(\delta_{ijj'}-\tau^2)\mathcal{X}_{ij}(\ttheta_0)\mathcal{X}_{ij'}(\ttheta_0)\trans\Bigg\},
\end{eqnarray}
where $\delta_{ijj^{'}}-\tau^2$ is aroused by $Cov\{\psi_\tau(e_{ij}),\psi_\tau(e_{ij^{'}})\}$, $\delta_{ijj^{'}}=P\left(e_{ij}<0,e_{ij^{'}}<0\right)$ measures  the dependence of residuals across different measurements from the same subject and $\ddelta=(\delta_{ijj^{'}};i=1,\cdots,N,j, j'=1,\cdots,n_i)\trans$.

\begin{thm}\label{thm2}
Under Assumptions (A1)-(A6) in the Web Appendix A, then under Model (\ref{eq:4}),  $\widehat{\ttheta}^{I}$ is $\sqrt{n}-$consistent and asymptotically normal, that is
$$
\sqrt{n}(\widehat{\ttheta}^{I}-\ttheta_0)\stackrel{d}{\longrightarrow}N\left\{{\bf 0},
\bH^{-1}\SSigma(\ddelta)
\bH^{-1}\right\},\quad \text{as $n\rightarrow\infty$.}
$$
where $\bH=\lim_{n\rightarrow\infty}\bH_n$ and $\SSigma(\ddelta)=\lim_{n\rightarrow\infty}\SSigma_n(\ddelta)$.
\end{thm}
Theorem \ref{thm2} establishes the asymptotic normality of the parameter estimators. For the purpose of statistical inference, both $f_{ij}(\cdot)$ and $\ddelta$ need to be estimated.
To estimate density $f_{ij}\{Q_Y(\tau;\ttheta_0|\bW_{ij})\}$ in $\bH$, we adopt the quotient estimation method of \cite{hendricks1992hierarchical}. The estimation of $\ddelta$ in $\SSigma(\ddelta)$ depends on the assumed correlation structure of $\bY_i$.

Next, we study theoretical properties of the GEE estimator.  Let $\bG_{il}=\mathcal{X}_{i}(\ttheta_0)\trans\bM_{li}\UUpsilon_i
\mathcal{X}_{i}(\ttheta_0)$,
where $\UUpsilon_i=diag(f_{ij}\{Q_Y(\tau;\ttheta_0|\bW_{ij})\},j=1,\cdots,n_i)$ and $\bG_N=(\bG_{1}\trans,\cdots,\bG_{v}\trans)\trans$, where $\bG_l=N^{-1}\sum_{i=1}^N\bG_{il}$ for $l=1,\cdots,v$.  Denote
$\XXi^0_{ll'}={N}^{-1}\sum_{i=1}^N\bS_{il}(\ttheta_0)\bS_{il'}(\ttheta_0)\trans
$
as the $(l,l')$th block of  $\XXi_N^0=\{\XXi_{ll'}^0\}_{l,l'=1}^v$.  The following theorem establishes the asymptotic normality of the GEE estimator.
\begin{thm}\label{thm:gee}
Suppose Assumptions (A1)-(A8) in the Web Appendix A hold, then under Model (\ref{eq:4}), there exists a local minimizer in (\ref{eq:eta}) such that $\widehat{\ttheta}^{C}$ is $\sqrt{N}$-consistent and asymptotically normal, that is
$$
\sqrt{N}(\widehat{\ttheta}^{C}-\ttheta_0)\stackrel{d}{\rightarrow}N
\left\{{\bf 0},(\bG\trans
\XXi^{-1}\bG)^{-1}
\right\}
$$
where $\bG=\lim_{N\rightarrow\infty}\bG_N$ and $\XXi=\lim_{N\rightarrow\infty}\XXi_N^0$.
\end{thm}

\subsection{Selection consistency}
Next, we establish the  selection consistency of the Schwarz-type quantile information criterion to estimate the number of kink points.
\begin{thm}\label{thm1}
Under the Assumptions (A1)-(A6) in the Web-Appendix A, we have that $P(\widehat{K}=K_0)\rightarrow1$ as $n\rightarrow\infty$.
\end{thm}
Theorem \ref{thm1} demonstrates that the estimated $\widehat{K}$ via minimizing the SIC  is equal to the true value $K_0$  with probability approaching 1 as the sample size $n$ goes to infinity. {In literature on change point detection such as \cite{fryzlewicz2014wild} and \cite{chan2014group}, the similar selection consistency for the number of change points has been also studied.}

\section{Testing the existence of kink points}
In this section, we focus on  testing whether there exists at least one kink point, without concerning the accurate number of kink points. To be specific, we are interested in the following null ($H_0$) and alternative ($H_1$) hypotheses,
\begin{equation}\label{eq:test}
\text{$H_0:$ $\beta_k=0$, for all $k=1,\cdots,K$. v.s. $H_1$: $\beta_k\neq0$ for some $k\in\{1,\cdots,K\}$.}
\end{equation}
To test (\ref{eq:test}), one may adopt the Wald-type test based on the asymptotic normality of the quantile estimator $\widehat{\ttheta}$. But it depends on the estimation of $K$. The use of likelihood ratio-based test is more complex as estimating $K$ is also needed and the limiting distribution is  much complicated. To avoid these problems, we turn to the score test based on the partial subgradient. Let $\xxi=(\alpha_0,\beta_0,\ggamma\trans)\trans$  and $\bU_{ij}=(1,X_{ij},\bZ_{ij}\trans)\trans$. We define
$$
R_n(t,\tau,\widehat{\xxi})=n^{-1/2}\sum_{i=1}^N\sum_{j=1}^{n_i}\psi_\tau(Y_{ij}-\widehat{\xxi}\trans\bU_{ij})(X_{ij}-t)I(X_{ij}\leq t),
$$
where $\widehat{\xxi}=\arg\min\sum_{i,j}\rho_\tau(Y_{ij}-\xxi\trans\bU_{ij}),$ which is the estimator without any kink point under the null hypothesis.
$R_n(t,\tau,\widehat{\xxi})$ can be regarded as the variant of the partial score of the quantile objective function with respect to $\beta_1$ at $\beta_1=0$  for the sub-sample with $X_{ij}$ below $t$.
The proposed score test statistic is
\begin{equation}\label{test1}
T_n(\tau)=\sup_{t\in\mathcal{T}}|R_n(t,\tau,\widehat{\xxi})|,
\end{equation}
where $\mathcal{T}$ is the compact set for the kink point $t$.
Intuitively, if the null hypothesis is true, the magnitudes of
$R_n(t,\tau,\widehat{\xxi})$ are relatively small which leads to the relative small value of $T_n(\tau)$; otherwise, the value of $T_n(\tau)$ is relatively large.
The score test (\ref{test1}) can be regarded as a CUSUM-type test. This test statistic is computationally appealing since $R_n(t,\tau,\widehat{\xxi})$  only requires estimating the model under the null hypothesis with no kink point so we avoid estimating either the number or the locations of kink points. As pointed by \cite{qu2008testing},  the score-type test statistic is also applicable for multiple change/kink points situation. 

To obtain the limiting distribution of $T_n(\tau)$, we introduce the following notations.
\begin{equation*}
\begin{array}{c}
\widehat{\bD}_n=n^{-1}\sum_{i,j}\bU_{ij}
\bU_{ij}\trans\widehat{f}_{ij}(\widehat{\xxi}\trans\bU_{ij}),\quad  \bD_n=n^{-1}\sum_{i,j}\bU_{ij}
\bU_{ij}\trans f_{ij}(\xxi\trans\bU_{ij}).
\end{array}
\end{equation*}

We provide the following limiting distribution of $T_n(\tau)$ under $H_0$.
\begin{thm}\label{thm:3}
Suppose that the Assumptions (A1)-(A6) and (A9) in the Web Appendix B hold. Under the null hypothesis $H_0$, we have
$$
T_n(\tau)\Rightarrow\sup_{t\in\mathcal{T}}|R(t)|,
$$
where $R(t)$ is a Gaussian process with mean zero and covariance function
\begin{eqnarray*}
&&\mathcal{W}(t,t^{'})=n^{-1}\Bigg[\sum_{i=1}^N\sum_{j=1}^{n_i}
\tau(1-\tau)\left\{(X_{ij}-t)I(X_{ij}\leq t)-\bD_1\trans(t)\bD^{-1}\bU_{ij}\right\}\times\\
&&\left\{(X_{ij}-{t}^{'})I(X_{ij}\leq {t}^{'})-\bD_1\trans({t}^{'})\bD^{-1}\bU_{ij}\right\}+\sum_{i=1}^N\sum_{j\neq j^{'}}(\delta_{ijj^{'}}-\tau^2)\left\{(X_{ij}-t)I(X_{ij}\leq t)\right.\\
&&\left.-\bD_1\trans(t)\bD^{-1}\bU_{ij}\right\}\left\{(X_{ij^{'}}-t^{'})I(X_{ij}\leq t^{'})-\bD_1\trans(t^{'})\bD^{-1}\bU_{ij^{'}}\right\} \Bigg]
\end{eqnarray*}
where $``\Rightarrow"$ denotes weak convergence, $\delta_{ijj^{'}}$ is defined in Theorem \ref{thm2} and  $\bD_1(t)=\lim_{n\rightarrow\infty}\bD_{n1}(t)$
with $\bD_{n1}(t)=n^{-1}\sum_{i,j}\bU_{ij}(X_{ij}-t)I(X_{ij}\leq t)f_{ij}(\xxi\trans\bU_{ij})$.
\end{thm}

\noindent Remark that the second part of the covariance function $\mathcal{W}(t,t^{'})$ captures the within-subject correlations of the longitudinal data, which makes it different from the independent data.

Next, we derive the limiting distribution of the test statistic under the following local alternative model in Theorem \ref{thm:4},
\begin{equation}\label{eq:alter}
Q_Y(\tau|\bW_{ij})=\alpha_0+\beta_0X_{ij}+n^{-1/2}\beta_1(X_{ij}-t)I(X_{ij}\geq t)+\ggamma\trans\bZ_{ij},\quad t\in\mathcal{T}.
\end{equation}

\begin{thm}\label{thm:4}
Suppose that the assumptions in Theorem \ref{thm:3} hold. Under the local alternative model (\ref{eq:alter}), we have
$$
T_n(\tau)\Rightarrow\sup_{t\in\mathcal{T}}|R(t)+q(t,\beta_1)|,
$$
where $R(t)$ is the same as in Theorem \ref{thm:3}, $q(t,\beta_1)=-\bD_1\trans(t)\bD^{-1}\bD_2(t,\beta_1)$ and $\bD_2(t,\beta_1)=\lim_{n\rightarrow\infty}\bD_{n2}(t,\beta_1)$
with
$\bD_{n2}(t,\beta_1)=n^{-1}\sum_{i,j}\bU_{ij}\beta_1(X_{ij}-t)I(X_{ij}>t)f_{ij}(\xxi\trans\bU_{ij})
$.
\end{thm}

However, the limiting null distribution of $T_n(\tau)$ takes the non-standard form and its critical values as well as the P-values can not be tabulated directly.  To calculate the P-values numerically, we first define an
asymptotic representation of $R_n(t)$ as
$$
R_n^*(t)=n^{-1/2}\sum_{i=1}^N\vartheta_{i}\sum_{j=1}^{n_i}\psi_{\tau}(Y_{ij}-\bU_{ij}\trans\widehat{\xxi})\left\{(X_{ij}-t)I(X_{ij}\leq t)-\widehat{\bD}_{n1}(t)\widehat{\bD}_n^{-1}\bU_{ij}\right\},
$$
where $\widehat{\bD}_{n1}(t)=n^{-1}\sum_{i,j}\bU_{ij}
(X_{ij}-t)I(X_{ij}\leq t)\widehat{f}_{ij}(\widehat{\xxi}\trans\bU_{ij})$ and
$\{\vartheta_{i};i=1,\cdots,N\}$ are a random sample from the standard normal distribution.
Theorem \ref{thm:5} shows the asymptotic representation   $R_n^*(t)$ shares the same limiting null distribution of $R_n(t)$.
\begin{thm}\label{thm:5}
Suppose that the assumptions in Theorem \ref{thm:3} hold,
under the null hypothesis, $R_n^*(t)$ converges to the Gaussian process $R(t)$ defined in Theorem \ref{thm:3}  as $n\rightarrow\infty$.
\end{thm}
Then, we develop a modified blockwise wild bootstrap method in Algorithm \ref{alg:1} based on $R_n^*(t)$ to characterize the limiting null distribution of $T_n(\tau)$.

\begin{algorithm}
\caption{~~A modified blockwise wild bootstrap method}\label{alg:1}
\begin{algorithmic}
\State {\bf Step 1.}
Calculate the test statistic $T_n(\tau)$ using the original data.
\State {\bf Step 2.}
Generate  $\{\vartheta_{i};i=1,\cdots,N\}$ from the standard normal distribution and calculate the bootstrapped test statistic $T_n^*(\tau)=\sup_{t\in\mathcal{T}}|R_n^*(t)|$.
\State{\bf Step 3.} Repeat Steps 1-2 B times to get $\{T_{nb}^*(\tau), b=1,\cdots,B\}$. The empirical P-value is the proportion of $T_{nb}(\tau)$'s exceeding $T_n(\tau)$ i.e.
$
\widehat{p}_n=B^{-1}\sum_{b=1}^BI\{T_{nb}^*(\tau)\geq T_n(\tau)\}.
$
\end{algorithmic}
\end{algorithm}

\section{Simulation Studies}
\subsection{Model descriptions}

In this section, {two data generation processes (DGPs) are considered   to  evaluate the finite sample performance of the proposed methods, including the asymptotic performance of two kinds of parameter estimators and the performance of the proposed score test procedure.
}

\noindent{\bf DGP 1:} We generate data from the following model,
\begin{equation}\label{eq:dgp}
\begin{array}{c}
Y_{ij}=\alpha_0+\alpha_1X_{ij}+\gamma Z_{ij}+\sum_{k=1}^K\beta_k(X_{ij}-t_k)I(X_{ij}>t_k)+e_{ij},
\end{array}
\end{equation}
where $i=1,\cdots,N, j=1,\cdots,n_i,$ $(\alpha_0,\alpha_1,\gamma)=(1,1,0.2)$ are fixed and $e_{ij}$'s are the error terms.
Similar to \cite{li2015quantile}, we consider three different cases: Case 1 (Compound Symmetry Correlation Structure), $e_{ij}=a_i+\epsilon_{ij}$, where $a_i\overset{i.i.d}{\sim}N(0,1)$ and $\epsilon_{ij}\overset{i.i.d}{\sim}t_3$; Case 2 (AR(1) Correlation Structure), $e_{ij}=v(X_{ij})u_{ij}$ where $v(x)=3.2-0.2x$, $u_{ij}=0.5u_{i,j-1}+\epsilon_{ij}$ and $\epsilon_{ij}\overset{i.i.d}{\sim}N(0,1)$ and Case 3 (Heteroscedastic Correlation Structure), $e_{ij}=a_i+g(X_{ij})\epsilon_{ij}$ where $g(x)=\sqrt{(3.2-0.2x)^2-1}$, $a_i\overset{i.i.d}{\sim}N(0,1)$ and $\epsilon_{ij}\overset{i.i.d}{\sim}N(0,1)$. In the Cases 1 and 3, $X_{ij}\overset{i.i.d}{\sim}U(0,10)$ and $Z_{ij}\overset{i.i.d}{\sim}U(0,10)$. In Case 2, $X_{i1}\overset{i.i.d}{\sim}U(0.5,7.5)$, $X_{ij}=X_{i,j-1}+0.5$ for $j>1$, and $Z_{ij}\overset{i.i.d}{\sim}U(0,10)$. For each case, we let $n_i\in\{6,7,8,9,10\}$ with equal size $N/5$ for each element to  add imbalance to the number of observations per subject. The size of total observations is $n=8N$. For all cases, the simulation was repeated 1000 times with the number of subjects $N=100,200, 400, 800$ and 1600, respectively.

{
\noindent{\bf DGP 2:} To evaluate the applicability of the proposed methods for the progesterone data analysis, we consider the similar model structure as in DGP 1 expect that  $X_{ij}$ is the day in cycles and $Z_{ij}$ is an indicator  taking 1 for the conceptive woman and 0 for non-conceptive woman in the progesterone data, see more details in Section 6. Let $(\alpha_0,\alpha_1,\gamma)=(-1,0,0.5)$ and the error term settings are the same as in DGP 1.}

\subsection{Selection consistency and parameter estimation}
We first check the selection consistency of $\widehat{K}$ in Theorem \ref{thm1}.  Three different kink effects are considered.  For DGP 1, we set (1) $K=1$ with $\beta_1=-2$ and $t_1=5$; (2) $K=2$ with $(\beta_1,\beta_2)=(-2,2)$ and $(t_1,t_2)=(3,6)$; (3) $K=3$ with $(\beta_1,\beta_2,\beta_3)=(-2,2,-2)$ and $(t_1,t_2,t_3)=(3,5,8)$. { For DGP 2, we set (1) $K=1$ with $\beta_1=0.5$ and $t_1=1$; (2) $K=2$ with $(\beta_1,\beta_2)=(0.5,-0.5)$ and ($t_1,t_2)=(-1,6)$; (3) $K=3$ with $(\beta_1,\beta_2,\beta_3)=(0.5,-0.5,0.5)$ and  $(t_1,t_2,t_3)=(-2,4,10)$. We consider quantile levels 0.25, 0.5 and 0.75, and repeat the simulation for each case 1000 times. Table \ref{tab:sim1} reports the percentages of the estimated $\widehat{K}$'s equaling to the true $K_0$ and the average running time across 1000 repetitions for  DGPs 1-2. We denote ``WI" and ``WC" as the estimator under the working independence framework and the GEE estimator via incorporating the working correlation structures, respectively.

We observe that as the number of  individuals  $N$ increases, the selection rates  gradually approach to 100\% for all cases across different $K$'s, which illustrates the selection consistency in Theorem \ref{thm1}. For DGP 1, it is obvious that the GEE estimator via incorporating the correlation information can effectively improve the selection rates across all cases with the price of much more computing time, and such phenomenon is even pronounced for $K=3$. However, for DGP 2, incorporating the correlation seems not to improve the selection rates. The possible reason is that the within-subject correlations may be weak in the progesterone data and the estimator under the working independence framework can still perform satisfactorily.}

\begin{table}
\centering
\footnotesize
\caption{The percentages of correctly selecting  $\widehat{K}=K$ ($K=1,2,3$) as well as their average running time (in second) of two different estimation strategies for DGPs 1-2. The results of DGP 1 are obtained across  sample sizes $N=200,400,800$ and 1600.}
\label{tab:sim1}
\scalebox{0.95}{
\begin{tabular}{lllllllllllllllll}
\hline
\hline
 &  & & \multicolumn{6}{c}{Correctly Selected Rate} & \multicolumn{6}{c}{Average Running Time (in second)}\\
\cmidrule(lr){4-9} \cmidrule(lr){10-15}
\multirow{2}{*}{$K$}& \multirow{2}{*}{$N$}& &  \multicolumn{3}{c}{WI} & \multicolumn{3}{c}{WC} & \multicolumn{3}{c}{WI}  & \multicolumn{3}{c}{WC}\\
\cmidrule(lr){4-6} \cmidrule(lr){7-9} \cmidrule(lr){10-12} \cmidrule(lr){13-15}
&  &  &  $0.25$ & 0.5 & 0.75 &   0.25 & 0.5 & 0.75  &  0.25 & 0.5 & 0.75 & 0.25 & 0.5 & 0.75\\
\hline
\multicolumn{15}{l}{DGP 1}\\
\hline
1	&	100	&	Case 1	&	0.978 	&	0.992 	&	0.982 	&	0.992 	&	0.994 	&	0.988 	&	1.134 	&	1.168 	&	1.270 	&	3.719 	&	3.428 	&	3.859 	\\
	&		&	Case 2	&	0.976 	&	0.988 	&	0.962 	&	0.996 	&	0.996 	&	0.988 	&	1.093 	&	1.108 	&	1.282 	&	3.812 	&	3.586 	&	4.059 	\\
	&		&	Case 3	&	0.976 	&	0.988 	&	0.972 	&	0.990 	&	0.994 	&	0.994 	&	1.117 	&	1.142 	&	1.255 	&	3.648 	&	3.395 	&	3.735 	\\
	&	200	&	Case 1	&	0.976 	&	0.994 	&	0.990 	&	0.992 	&	1.000 	&	0.994 	&	1.822 	&	1.999 	&	2.174 	&	6.460 	&	6.025 	&	6.915 	\\
	&		&	Case 2	&	0.974 	&	0.992 	&	0.984 	&	0.990 	&	0.994 	&	0.998 	&	1.741 	&	1.899 	&	2.045 	&	6.561 	&	6.340 	&	7.062 	\\
	&		&	Case 3	&	0.970 	&	0.988 	&	0.968 	&	0.982 	&	0.996 	&	0.984 	&	1.786 	&	1.959 	&	2.179 	&	6.250 	&	5.947 	&	6.801 	\\
	&	400	&	Case 1	&	0.996 	&	0.994 	&	0.998 	&	0.998 	&	0.998 	&	1.000 	&	3.304 	&	4.085 	&	4.579 	&	11.198 	&	10.891 	&	12.483 	\\
	&		&	Case 2	&	0.986 	&	0.990 	&	0.970 	&	0.994 	&	0.996 	&	0.988 	&	3.171 	&	3.818 	&	4.444 	&	10.915 	&	11.022 	&	12.471 	\\
	&		&	Case 3	&	0.990 	&	0.994 	&	0.990 	&	0.994 	&	1.000 	&	0.994 	&	3.335 	&	3.919 	&	4.672 	&	10.484 	&	10.186 	&	11.991 	\\
	&	800	&	Case 1	&	0.990 	&	1.000 	&	0.986 	&	0.996 	&	1.000 	&	0.998 	&	8.136 	&	10.861 	&	12.945 	&	22.083 	&	22.716 	&	26.511 	\\
	&		&	Case 2	&	0.990 	&	0.998 	&	0.982 	&	0.996 	&	1.000 	&	0.990 	&	7.378 	&	9.776 	&	11.673 	&	21.739 	&	22.741 	&	26.576 	\\
	&		&	Case 3	&	0.984 	&	0.998 	&	0.996 	&	0.990 	&	1.000 	&	1.000 	&	8.109 	&	10.330 	&	12.155 	&	20.776 	&	21.949 	&	25.310 	\\
	&	1600	&	Case 1	&	0.992 	&	0.990 	&	0.990 	&	0.996 	&	0.992 	&	0.994 	&	22.635 	&	32.013 	&	37.940 	&	47.065 	&	53.765 	&	62.473 	\\
	&		&	Case 2	&	0.990 	&	0.992 	&	0.986 	&	0.996 	&	0.992 	&	0.988 	&	18.679 	&	26.835 	&	33.049 	&	43.580 	&	50.221 	&	59.142 	\\
	&		&	Case 3	&	0.982 	&	0.992 	&	0.988 	&	0.988 	&	0.992 	&	0.990 	&	17.546 	&	24.537 	&	29.091 	&	34.695 	&	39.894 	&	46.472 	\\
\hline
2	&	100	&	Case 1	&	0.964 	&	0.976 	&	0.968 	&	0.986 	&	0.980 	&	0.980 	&	1.981 	&	2.009 	&	2.153 	&	4.033 	&	3.848 	&	4.223 	\\
	&		&	Case 2	&	0.944 	&	0.962 	&	0.952 	&	0.948 	&	0.972 	&	0.958 	&	1.878 	&	1.939 	&	1.960 	&	4.114 	&	3.970 	&	4.129 	\\
	&		&	Case 3	&	0.966 	&	0.986 	&	0.982 	&	0.980 	&	0.988 	&	0.988 	&	1.968 	&	1.918 	&	2.102 	&	3.973 	&	3.757 	&	4.114 	\\
	&	200	&	Case 1	&	0.966 	&	0.976 	&	0.960 	&	0.988 	&	0.980 	&	0.980 	&	3.288 	&	3.561 	&	3.983 	&	7.178 	&	7.042 	&	7.834 	\\
	&		&	Case 2	&	0.980 	&	0.986 	&	0.990 	&	0.986 	&	0.990 	&	0.994 	&	3.029 	&	3.314 	&	3.654 	&	7.336 	&	7.174 	&	7.954 	\\
	&		&	Case 3	&	0.972 	&	0.972 	&	0.978 	&	0.972 	&	0.973 	&	0.982 	&	3.218 	&	3.609 	&	3.840 	&	7.171 	&	7.064 	&	7.706 	\\
	&	400	&	Case 1	&	0.978 	&	0.978 	&	0.966 	&	0.982 	&	0.980 	&	0.970 	&	6.340 	&	7.693 	&	8.668 	&	13.559 	&	14.111 	&	15.959 	\\
	&		&	Case 2	&	0.986 	&	0.982 	&	0.978 	&	0.988 	&	0.986 	&	0.980 	&	5.478 	&	6.538 	&	7.678 	&	13.225 	&	13.439 	&	15.406 	\\
	&		&	Case 3	&	0.970 	&	0.978 	&	0.968 	&	0.972 	&	0.984 	&	0.974 	&	5.871 	&	6.804 	&	7.936 	&	12.662 	&	12.697 	&	14.679 	\\
	&	800	&	Case 1	&	0.970 	&	0.972 	&	0.978 	&	0.976 	&	0.974 	&	0.978 	&	14.112 	&	19.039 	&	22.015 	&	26.742 	&	30.280 	&	34.640 	\\
	&		&	Case 2	&	0.966 	&	0.990 	&	0.970 	&	0.968 	&	0.992 	&	0.976 	&	12.719 	&	16.223 	&	20.050 	&	27.079 	&	29.265 	&	34.196 	\\
	&		&	Case 3	&	0.986 	&	0.974 	&	0.970 	&	0.992 	&	0.978 	&	0.976 	&	13.529 	&	18.367 	&	21.998 	&	26.285 	&	29.776 	&	34.702 	\\
	&	1600	&	Case 1	&	0.986 	&	0.988 	&	0.966 	&	0.986 	&	0.988 	&	0.968 	&	37.306 	&	54.870 	&	65.828 	&	60.859 	&	75.764 	&	88.928 	\\
	&		&	Case 2	&	0.972 	&	0.990 	&	0.974 	&	0.972 	&	0.990 	&	0.980 	&	29.898 	&	43.507 	&	54.844 	&	55.356 	&	66.385 	&	79.601 	\\
	&		&	Case 3	&	0.970 	&	0.984 	&	0.972 	&	0.970 	&	0.984 	&	0.972 	&	33.575 	&	49.053 	&	60.632 	&	55.736 	&	69.157 	&	82.570 	\\
\hline
3	&	100	&	Case 1	&	0.892 	&	0.914 	&	0.879 	&	0.898 	&	0.926 	&	0.897 	&	3.033 	&	3.142 	&	3.467 	&	4.316 	&	4.374 	&	4.816 	\\
	&		&	Case 2	&	0.834 	&	0.908 	&	0.866 	&	0.886 	&	0.911 	&	0.884 	&	2.881 	&	3.183 	&	3.131 	&	4.268 	&	4.543 	&	4.578 	\\
	&		&	Case 3	&	0.875 	&	0.916 	&	0.887 	&	0.852 	&	0.926 	&	0.892 	&	3.142 	&	3.118 	&	3.407 	&	4.377 	&	4.330 	&	4.601 	\\
	&	200	&	Case 1	&	0.922 	&	0.934 	&	0.920 	&	0.937 	&	0.940 	&	0.922 	&	4.834 	&	5.873 	&	6.287 	&	7.189 	&	7.988 	&	8.607 	\\
	&		&	Case 2	&	0.916 	&	0.926 	&	0.915 	&	0.921 	&	0.932 	&	0.916 	&	4.583 	&	5.030 	&	6.014 	&	7.170 	&	7.457 	&	8.480 	\\
	&		&	Case 3	&	0.937 	&	0.948 	&	0.928 	&	0.941 	&	0.951 	&	0.931 	&	4.911 	&	5.151 	&	6.138 	&	7.125 	&	7.242 	&	8.439 	\\
	&	400	&	Case 1	&	0.955 	&	0.964 	&	0.956 	&	0.958 	&	0.968 	&	0.962 	&	9.610 	&	11.827 	&	13.326 	&	13.638 	&	15.656 	&	17.508 	\\
	&		&	Case 2	&	0.929 	&	0.955 	&	0.948 	&	0.931 	&	0.959 	&	0.933 	&	8.558 	&	10.751 	&	11.617 	&	12.996 	&	15.174 	&	15.894 	\\
	&		&	Case 3	&	0.932 	&	0.957 	&	0.951 	&	0.935 	&	0.964 	&	0.957 	&	8.854 	&	11.039 	&	12.605 	&	12.769 	&	14.830 	&	16.607 	\\
	&	800	&	Case 1	&	0.958 	&	0.976 	&	0.976 	&	0.960 	&	0.983 	&	0.978 	&	20.928 	&	28.144 	&	33.776 	&	28.268 	&	34.774 	&	41.035 	\\
	&		&	Case 2	&	0.946 	&	0.971 	&	0.958 	&	0.952 	&	0.975 	&	0.964 	&	17.059 	&	24.043 	&	30.399 	&	24.611 	&	31.537 	&	38.972 	\\
	&		&	Case 3	&	0.964 	&	0.984 	&	0.967 	&	0.970 	&	0.988 	&	0.973 	&	18.152 	&	26.441 	&	31.027 	&	25.054 	&	33.069 	&	38.561 	\\
	&	1600	&	Case 1	&	0.986 	&	0.984 	&	0.988 	&	0.986 	&	0.984 	&	0.990 	&	58.955 	&	86.495 	&	93.822 	&	72.267 	&	99.062 	&	99.383 	\\
	&		&	Case 2	&	0.975 	&	0.979 	&	0.964 	&	0.976 	&	0.979 	&	0.968 	&	47.195 	&	70.414 	&	93.301 	&	61.775 	&	84.341 	&	98.438 	\\
	&		&	Case 3	&	0.973 	&	0.975 	&	0.968 	&	0.973 	&	0.975 	&	0.968 	&	49.381 	&	78.154 	&	95.104 	&	61.952 	&	90.252 	&	97.583 	\\
\hline
\multicolumn{15}{l}{DGP 2}\\
\hline
1	&	&	Case 1	&	0.996	&	1.000	&	0.996	&	0.990	&	0.999	&	0.970	&	1.260	&	1.239	&	1.304	&	3.118	&	2.658	&	3.202	\\
	&	&	Case 2	&	0.956	&	0.983	&	0.944	&	0.956	&	0.982	&	0.928	&	1.448	&	1.350	&	1.513	&	3.631	&	3.100	&	3.768	\\
	&	&	Case 3	&	0.996	&	0.998	&	0.994	&	0.955	&	0.998	&	0.970	&	1.281	&	1.239	&	1.335	&	3.184	&	2.668	&	3.248	\\
2	&	&	Case 1	&	0.998	&	1.000	&	0.998	&	0.977	&	0.996	&	0.956	&	2.854	&	2.769	&	2.769	&	5.441	&	4.705	&	5.414	\\
	&	&	Case 2	&	0.988	&	0.992	&	0.984	&	0.924	&	0.994	&	0.901	&	3.060	&	3.004	&	2.989	&	6.333	&	5.309	&	6.415	\\
	&	&	Case 3	&	0.998	&	0.999	&	1.000	&	0.960	&	0.999	&	0.958	&	2.912	&	2.819	&	2.866	&	5.613	&	4.759	&	5.576	\\
3	&	&	Case 1	&	0.981	&	0.986	&	0.968	&	0.647	&	0.969	&	0.476	&	6.581	&	6.231	&	6.465	&	8.439	&	9.038	&	7.238	\\
	&	&	Case 2	&	0.986	&	0.998	&	0.963	&	0.655	&	0.991	&	0.393	&	6.290	&	6.095	&	6.054	&	8.390	&	9.122	&	6.550	\\
	&	&	Case 3	&	0.976	&	0.982	&	0.967	&	0.702	&	0.961	&	0.484	&	6.049	&	5.811	&	5.988	&	8.091	&	8.394	&	6.671	\\
\hline
\end{tabular}}
\end{table}

Next, we  evaluate the finite sample performance of two proposed methods as well as the least squared (LS) estimator. For saving space, we only report the parameter estimation results of $K=2$ for DGP 1 when $N=400$  and DGP 2 for all error cases in Table \ref{tab:sim2}. The remaining results for DGP 1 based on $N=200,800$ and 1600 for each case  are given in the Web Appendix C in the supplement.   As shown in the Table,  all the parameter estimates have sufficiently small biases and the biases diminish to zero as the sample size increases, which demonstrates the consistency of the estimators. Besides, the Monte Carlo standard deviations (SD) are quite close to the estimated standard error (ESE) especially as  $N$ increases, which validates that the derived limiting distribution provides reasonable asymptotic variances.
The coverage rates of 95\% Wald confidence intervals (CovP) for the regression coefficients $(\alpha_0,\alpha_1,\gamma,\beta_1,\beta_2)$ are very close to the nominal level 95\%, while those for kink points $(t_1,t_2)$ are relatively indecent. But when the number of individuals increases to $N=1600$, their CovPs also becomes the nominal level. More discussions on the  CovPs for kink points are included in the Appendix C.
In summary, the methods based on quantile regression can generally perform better than the least squared estimators especially for heteroscedastic errors. The GEE estimators via incorporating the working correlations within subjects
are generally more efficient than the estimators under the working independence framework.

\begin{table}
\centering
\footnotesize
\caption{Simulation results of two proposed methods (WI and WC) and the LS method across 1000 repetitions for DGP 1 with $N=400$ and DGP 2 when $K=2$.}
\label{tab:sim2}
\scalebox{0.9}
{\begin{tabular}{cccccccccccccccc}
\hline
\hline
 &\multicolumn{5}{c}{WI} & \multicolumn{5}{c}{WC} & \multicolumn{5}{c}{LS}\\
\cmidrule(lr){2-6} \cmidrule(lr){7-11} \cmidrule(lr){12-16}
&		Bias	&	SD	&	ESE	&	MSE	&	CovP	&	Bias	&	SD	&	ESE	&	MSE	&	CovP	&	Bias	&	SD	&	ESE	&	MSE	&	CovP	\\

\hline
\multicolumn{16}{l}{DGP 1, Case 1}\\
\hline
$\alpha_0$	&	0.010 	&	0.184 	&	0.183 	&	0.034 	&	0.956 	&	0.005 	&	0.147 	&	0.152 	&	0.021 	&	0.944 	&	0.009 	&	0.172 	&	0.173 	&	0.029 	&	0.956 	\\
$\alpha_1$	&	0.001 	&	0.014 	&	0.014 	&	0.000 	&	0.960 	&	0.001 	&	0.011 	&	0.012 	&	0.000 	&	0.970 	&	0.000 	&	0.014 	&	0.014 	&	0.000 	&	0.950 	\\
$\beta_1$	&	-0.004 	&	0.094 	&	0.089 	&	0.009 	&	0.924 	&	-0.002 	&	0.070 	&	0.070 	&	0.005 	&	0.948 	&	-0.003 	&	0.082 	&	0.083 	&	0.007 	&	0.942 	\\
$\beta_2$	&	-0.009 	&	0.132 	&	0.127 	&	0.018 	&	0.942 	&	-0.003 	&	0.098 	&	0.100 	&	0.010 	&	0.946 	&	-0.013 	&	0.125 	&	0.118 	&	0.016 	&	0.930 	\\
$\gamma$	&	0.020 	&	0.111 	&	0.107 	&	0.013 	&	0.954 	&	0.007 	&	0.085 	&	0.085 	&	0.007 	&	0.948 	&	0.020 	&	0.103 	&	0.100 	&	0.011 	&	0.938 	\\
$t_1$	&	0.012 	&	0.127 	&	0.109 	&	0.016 	&	0.908 	&	0.005 	&	0.094 	&	0.087 	&	0.009 	&	0.926 	&	0.013 	&	0.110 	&	0.101 	&	0.012 	&	0.910 	\\
$t_2$	&	0.000 	&	0.121 	&	0.101 	&	0.015 	&	0.898 	&	-0.001 	&	0.090 	&	0.081 	&	0.008 	&	0.918 	&	-0.008 	&	0.117 	&	0.094 	&	0.014 	&	0.896 	\\

\hline
\multicolumn{16}{l}{DGP 1, Case 2}\\
\hline
$\alpha_0$	&	0.009 	&	0.346 	&	0.356 	&	0.119 	&	0.950 	&	0.008 	&	0.296 	&	0.284 	&	0.087 	&	0.924 	&	0.002 	&	0.294 	&	0.300 	&	0.087 	&	0.960 	\\
$\alpha_1$	&	0.000 	&	0.020 	&	0.020 	&	0.000 	&	0.948 	&	0.000 	&	0.016 	&	0.016 	&	0.000 	&	0.946 	&	0.000 	&	0.018 	&	0.018 	&	0.000 	&	0.952 	\\
$\beta_1$	&	0.007 	&	0.194 	&	0.177 	&	0.038 	&	0.930 	&	0.001 	&	0.152 	&	0.136 	&	0.023 	&	0.908 	&	0.008 	&	0.154 	&	0.143 	&	0.024 	&	0.936 	\\
$\beta_2$	&	-0.042 	&	0.224 	&	0.230 	&	0.052 	&	0.962 	&	-0.026 	&	0.178 	&	0.178 	&	0.032 	&	0.950 	&	-0.033 	&	0.179 	&	0.184 	&	0.033 	&	0.958 	\\
$\gamma$	&	0.040 	&	0.167 	&	0.157 	&	0.029 	&	0.940 	&	0.027 	&	0.124 	&	0.123 	&	0.016 	&	0.952 	&	0.030 	&	0.132 	&	0.127 	&	0.018 	&	0.946 	\\
$t_1$	&	0.018 	&	0.236 	&	0.192 	&	0.056 	&	0.876 	&	0.015 	&	0.175 	&	0.151 	&	0.031 	&	0.898 	&	0.007 	&	0.187 	&	0.155 	&	0.035 	&	0.882 	\\
$t_2$	&	-0.013 	&	0.158 	&	0.138 	&	0.025 	&	0.910 	&	-0.016 	&	0.117 	&	0.108 	&	0.014 	&	0.904 	&	-0.011 	&	0.126 	&	0.112 	&	0.016 	&	0.908 	\\
\hline
\multicolumn{16}{l}{DGP 1, Case 3}\\
\hline
$\alpha_0$	&	-0.023 	&	0.287 	&	0.281 	&	0.083 	&	0.936 	&	-0.009 	&	0.241 	&	0.230 	&	0.058 	&	0.948 	&	-0.021 	&	0.266 	&	0.258 	&	0.071 	&	0.936 	\\
$\alpha_1$	&	0.000 	&	0.018 	&	0.018 	&	0.000 	&	0.956 	&	-0.001 	&	0.015 	&	0.015 	&	0.000 	&	0.960 	&	0.001 	&	0.015 	&	0.016 	&	0.000 	&	0.964 	\\
$\beta_1$	&	0.010 	&	0.152 	&	0.143 	&	0.023 	&	0.938 	&	0.007 	&	0.124 	&	0.114 	&	0.015 	&	0.932 	&	0.007 	&	0.140 	&	0.131 	&	0.019 	&	0.924 	\\
$\beta_2$	&	-0.030 	&	0.190 	&	0.189 	&	0.037 	&	0.956 	&	-0.015 	&	0.150 	&	0.149 	&	0.023 	&	0.946 	&	-0.019 	&	0.168 	&	0.170 	&	0.028 	&	0.942 	\\
$\gamma$	&	0.021 	&	0.146 	&	0.133 	&	0.022 	&	0.932 	&	0.008 	&	0.112 	&	0.105 	&	0.012 	&	0.940 	&	0.011 	&	0.122 	&	0.116 	&	0.015 	&	0.942 	\\
$t_1$	&	0.015 	&	0.190 	&	0.160 	&	0.036 	&	0.922 	&	0.004 	&	0.150 	&	0.127 	&	0.023 	&	0.906 	&	0.009 	&	0.168 	&	0.144 	&	0.028 	&	0.914 	\\
$t_2$	&	-0.001 	&	0.140 	&	0.120 	&	0.020 	&	0.892 	&	0.001 	&	0.104 	&	0.096 	&	0.011 	&	0.938 	&	-0.003 	&	0.116 	&	0.103 	&	0.013 	&	0.896 	\\
\hline
\multicolumn{16}{l}{DGP 2, Case 1}\\
\hline
$\alpha_0$	&	-0.034 	&	0.217 	&	0.204 	&	0.048 	&	0.926 	&	-0.023 	&	0.202 	&	0.193 	&	0.044 	&	0.906 	&	-0.023 	&	0.212 	&	0.199 	&	0.046 	&	0.936 	\\
$\alpha_1$	&	0.001 	&	0.282 	&	0.236 	&	0.079 	&	0.901 	&	0.007 	&	0.263 	&	0.262 	&	0.069 	&	0.952 	&	0.006 	&	0.263 	&	0.256 	&	0.069 	&	0.938 	\\
$\beta_1$	&	-0.005 	&	0.032 	&	0.031 	&	0.001 	&	0.932 	&	-0.004 	&	0.028 	&	0.028 	&	0.001 	&	0.928 	&	-0.003 	&	0.032 	&	0.030 	&	0.001 	&	0.946 	\\
$\beta_2$	&	0.010 	&	0.048 	&	0.048 	&	0.002 	&	0.934 	&	0.009 	&	0.042 	&	0.043 	&	0.002 	&	0.950 	&	0.011 	&	0.050 	&	0.048 	&	0.003 	&	0.940 	\\
$\gamma$	&	-0.009 	&	0.046 	&	0.045 	&	0.002 	&	0.944 	&	-0.007 	&	0.039 	&	0.040 	&	0.002 	&	0.952 	&	-0.011 	&	0.048 	&	0.045 	&	0.002 	&	0.934 	\\
$t_1$	&	-0.007 	&	0.406 	&	0.385 	&	0.164 	&	0.940 	&	0.006 	&	0.365 	&	0.349 	&	0.133 	&	0.924 	&	0.032 	&	0.412 	&	0.377 	&	0.170 	&	0.942 	\\
$t_2$	&	0.008 	&	0.392 	&	0.376 	&	0.153 	&	0.938 	&	0.002 	&	0.341 	&	0.342 	&	0.116 	&	0.940 	&	-0.016 	&	0.384 	&	0.373 	&	0.147 	&	0.948 	\\		
\hline
\multicolumn{16}{l}{DGP 2, Case 2}\\
\hline
	$\alpha_0$	&	-0.027 	&	0.228 	&	0.222 	&	0.053 	&	0.938 	&	-0.021 	&	0.219 	&	0.192 	&	0.049 	&	0.922 	&	-0.036 	&	0.253 	&	0.251 	&	0.065 	&	0.944 	\\
$\alpha_1$	&	-0.004 	&	0.164 	&	0.156 	&	0.027 	&	0.936 	&	-0.006 	&	0.166 	&	0.141 	&	0.028 	&	0.910 	&	-0.005 	&	0.193 	&	0.189 	&	0.037 	&	0.940 	\\
$\beta_1$	&	-0.004 	&	0.039 	&	0.039 	&	0.002 	&	0.938 	&	-0.003 	&	0.037 	&	0.032 	&	0.001 	&	0.928 	&	-0.006 	&	0.046 	&	0.045 	&	0.002 	&	0.934 	\\
$\beta_2$	&	0.018 	&	0.066 	&	0.065 	&	0.005 	&	0.936 	&	0.012 	&	0.063 	&	0.057 	&	0.004 	&	0.916 	&	0.025 	&	0.082 	&	0.076 	&	0.007 	&	0.928 	\\
$\gamma$	&	-0.019 	&	0.065 	&	0.063 	&	0.005 	&	0.930 	&	-0.013 	&	0.063 	&	0.056 	&	0.004 	&	0.912 	&	-0.026 	&	0.077 	&	0.073 	&	0.007 	&	0.940 	\\
$t_1$	&	0.048 	&	0.527 	&	0.462 	&	0.280 	&	0.914 	&	0.026 	&	0.486 	&	0.385 	&	0.236 	&	0.901 	&	0.063 	&	0.546 	&	0.497 	&	0.302 	&	0.914 	\\
$t_2$	&	-0.012 	&	0.485 	&	0.455 	&	0.235 	&	0.920 	&	0.009 	&	0.460 	&	0.383 	&	0.211 	&	0.910 	&	-0.033 	&	0.656 	&	0.503 	&	0.430 	&	0.876 	\\	
\hline
\multicolumn{16}{l}{DGP 2, Case 3}\\
\hline
	$\alpha_0$	&	-0.036 	&	0.220 	&	0.205 	&	0.050 	&	0.922 	&	-0.029 	&	0.211 	&	0.185 	&	0.046 	&	0.908 	&	-0.040 	&	0.232 	&	0.209 	&	0.055 	&	0.928 	\\
$\alpha_1$	&	0.034 	&	0.270 	&	0.265 	&	0.074 	&	0.940 	&	0.029 	&	0.273 	&	0.236 	&	0.075 	&	0.916 	&	0.024 	&	0.266 	&	0.262 	&	0.071 	&	0.936 	\\
$\beta_1$	&	-0.003 	&	0.032 	&	0.032 	&	0.001 	&	0.936 	&	-0.003 	&	0.029 	&	0.029 	&	0.001 	&	0.942 	&	-0.005 	&	0.035 	&	0.033 	&	0.001 	&	0.938 	\\
$\beta_2$	&	0.015 	&	0.053 	&	0.050 	&	0.003 	&	0.952 	&	0.011 	&	0.047 	&	0.045 	&	0.002 	&	0.958 	&	0.016 	&	0.052 	&	0.050 	&	0.003 	&	0.950 	\\
$\gamma$	&	-0.015 	&	0.048 	&	0.047 	&	0.003 	&	0.944 	&	-0.011 	&	0.042 	&	0.042 	&	0.002 	&	0.948 	&	-0.016 	&	0.052 	&	0.047 	&	0.003 	&	0.920 	\\
$t_1$	&	0.038 	&	0.405 	&	0.395 	&	0.165 	&	0.938 	&	0.027 	&	0.376 	&	0.361 	&	0.142 	&	0.932 	&	0.025 	&	0.464 	&	0.402 	&	0.215 	&	0.908 	\\
$t_2$	&	-0.028 	&	0.409 	&	0.386 	&	0.168 	&	0.932 	&	-0.028 	&	0.365 	&	0.353 	&	0.134 	&	0.948 	&	-0.025 	&	0.437 	&	0.388 	&	0.192 	&	0.940 	\\
\hline
\end{tabular}}
\begin{tablenotes}
   \footnotesize
   \item  Note: Bias is empirical bias; SD is the Monte Carlo standard deviation; ESE is the estimated standard error; MSE is the mean square error and  CovP is the coverage probability of 95\% Wald confidence interval.
    \end{tablenotes}
\end{table}

{We remark that the asymptotic variances of the WI estimators under the working independence framework
are based on the correlation structure parameter $\ddelta$. Following \cite{li2015quantile}, we adopt a general  form to obtain  the  covariance matrix estimation for $\SSigma(\ddelta)$ in Theorem \ref{thm2}. To be specific,  $\delta_{ijj'}-\tau^2$ here can be estimated by
$I(\hat{e}_{ij}<0,\hat{e}_{ij'}<0)-2^{-1}\tau
\left\{I(\hat{e}_{ij}<0)+I(\hat{e}_{ij'}<0) \right\},$
where $\hat{e}_{ij}$'s  are estimated residuals. In real data analysis, we also adopt this general form to approximately  depict the within-subject correlations.
For the GEE estimator using the QIF approach, we choose  three basis matrices $\bM_{1i}$, $\bM_{2i}$ and $\bM_{3i}$ as specified in Section 2.2.2 to incorporate within-subject correlation.}

\subsection{Power analysis}
Next, we evaluate the performance of the proposed  testing procedure for the existence of the kink points using simulations. We consider varying kink effects with $K=1$ and 2 kink points and other parameters remain unchanged as DGPs 1-2. For DGP 1, we let $\beta_1=0,0.1,0.3,0.5,0.7$ for $K=1$ and $\beta_1=-\beta_2=0,0.1,0.3,0.5,0.7$ for $K=2$, and for DGP 2, $\beta_1=0,0.03,0.05,0.08,0.1$ for $K=1$ and $\beta_1=-\beta_2=0,0.1,0.2,0.3,0.4$ for $K=2$. For each scenario, we conduct 10000 repetitions for the type I error i.e. when there is no kink point and 1000 repetitions for the power when kink points exist. The P-values of the quantile score test are computed by using 300 bootstrap replicates i.e. $B=300$ in Algorithm \ref{alg:1}.

{The results including the average running time are summarized in Tables 8-9 in the Appendix D of the supplement. Figure \ref{fig:1} depicts the empirical power curves across $\tau= 0.1,0.25,0.5,0.75$ and 0.9 for  Cases 1-3 with $K=1$ kink point and the results for $K=2$ is showed in  the Web Appendix D.}  As shown in the Figure, when $\beta_1 = 0$, i.e. under the null hypothesis, the type I errors are reasonably close to the nominal level 5\%, which suggests that our proposed quantile score test can control the type I errors. As $\beta_1$  increases, i.e. the kink effect gets enhanced, the local powers gradually approach to one for all quantiles. We also observe that the powers at the tail quantile levels (e.g. $\tau= 0.1$ and 0.9) approach at a slower rate than that at the moderate quantile levels such as $\tau = 0.5$. It is common due to the asymmetry of observations at extreme quantiles and can be alleviated by increasing the sample size. Therefore, we conclude that the score-based test can also identify the existence of kink points well.

 \begin{figure}[!h]
\centering
\includegraphics[scale=0.85]{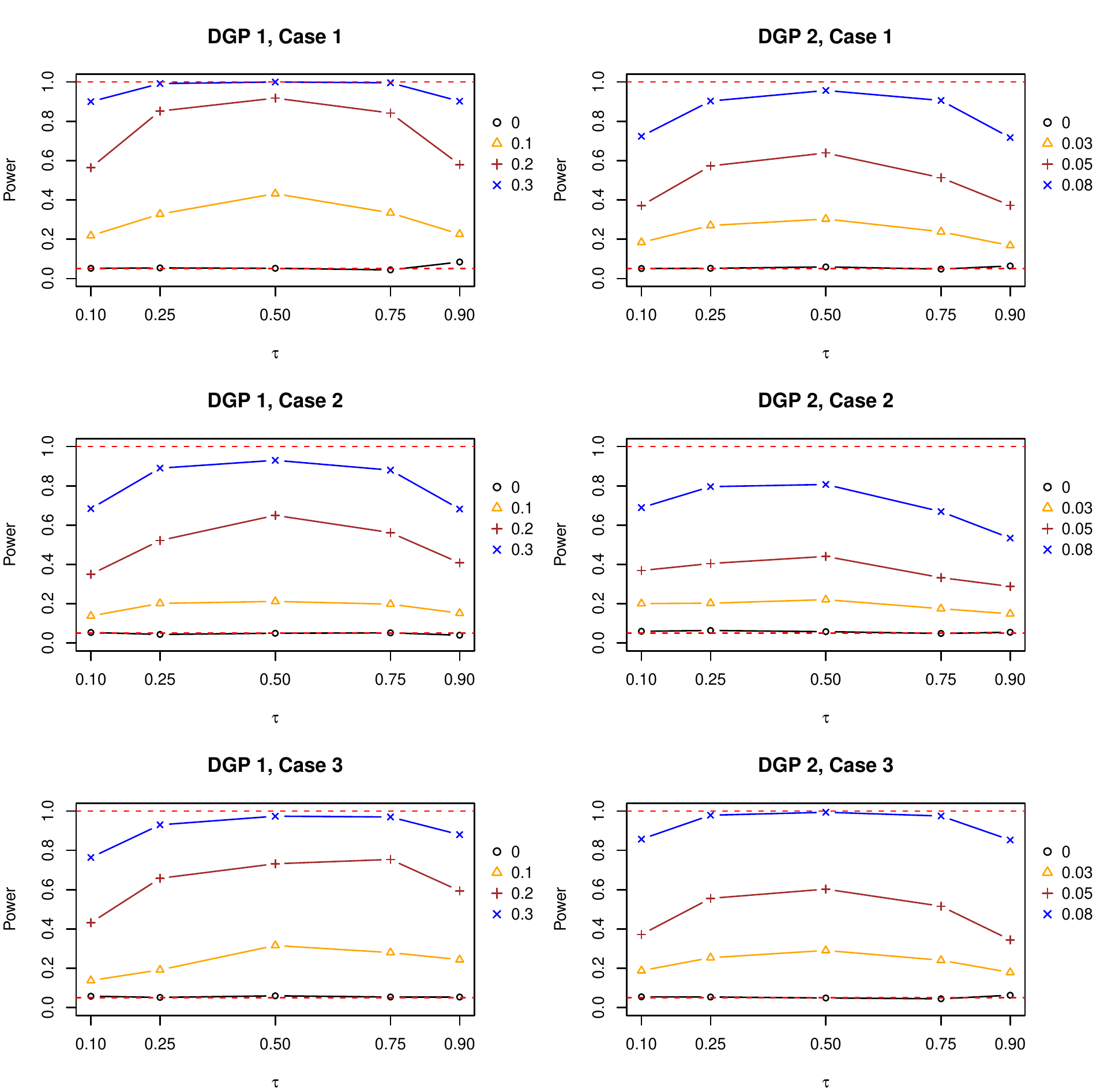}
\caption[]{\label{fig:1}
The local empirical powers when $K=1$ for DGP 1 with $N=400$ and DGP 2 across $\tau=0.10, 0.25, 0.50,0.75,0.90$. We conduct 10000 repetitions for the Type I errors and 1000 repetitions for the local powers.  The bottom and upside red dashed lines represent the nominal level 0.05 and 1, respectively.}
\end{figure}

\section{Progesterone Data Analysis}
In this section, we analyze the longitudinal progesterone data in \cite{munro1991relationship}. This longitudinal research followed  a total of 51 women  with healthy reproductive function enrolled in an artificial insemination clinic. The data set consists of two groups:  conceptive  and  non-conceptive. In the non-conceptive group, the urinary metabolite progesterone was measured for one to five menstrual cycles for a woman while in the conceptive group, each woman only contributed for one menstrual cycle. There are  22 conceptive and 69 non-conceptive women' menstrual cycles. At each cycle, the measurements were  recorded per day from 8 days before the day of ovulation and until 15 days after the ovulation, totally 24 days in a cycle. To make the progesterone curves have the same design point, the data had been aligned and truncated by the day of ovulation (Day 0). However, due to some reason, not all measurements in a cycle are fully observed, which results in some missing values.   After removing these   values, we get  a sample size of 2004 observations  among  $N=91$ women's menstrual cycles with each cycles contributing from 9 to 24 observations.

\begin{figure}[!h]
\centering
\includegraphics[scale=0.8]{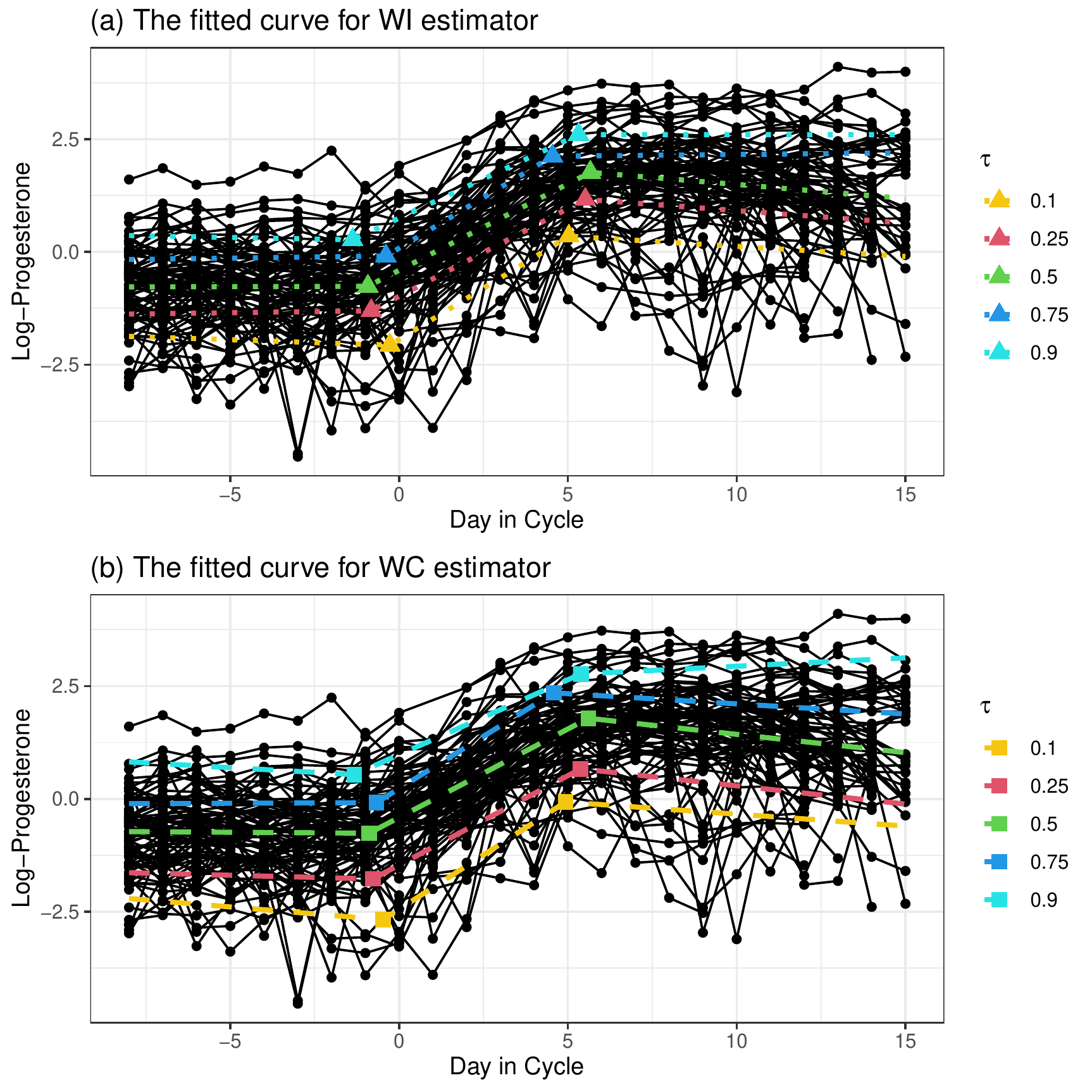}
\caption[]{\label{fig:2}
Scatter plots between the logarithm of progesterone and the day in cycle  with the fitted  curves for (a) the estimator under the working independence framework (WI) and (b) the GEE estimator incorporating the working correlations (WC)
at quantile levels 0.1, 0.25, 0.5, 0.75 and 0.9. $\blacktriangle$ and $\blacksquare$ denotes the estimated kink points using different methods.}
\end{figure}

Several researchers includes \cite{brumback1998smoothing}, \cite{fan2000two} and \cite{wu2002local} have analyzed this data set by using nonparametric regression methods. All these researches suggested that there exists nonlinear effect between the logarithmic progesterone and the day in cycles.
Figure \ref{fig:2}  presents a plot for the 91 progesterone curves. The logarithmic progesterone remains rather stable at first 8 days, but rises sharply after ovulation. The uptrend  persists several days and then drops slightly.  It seems that the progesterone curves experience two structural changes but the number remains to be identified. To validate this observation and identify the locations of kink points,  we re-analyze the data using the proposed  MKQR model for the longitudinal data,
\begin{equation}\label{eq:real}
Y_{ij}=\alpha_0+\alpha_1X_{ij}+\sum_{k=1}^K\beta_k(X_{ij}-t_k)_++\gamma Z_{ij}+e_{ij},
\end{equation}
where $Y_{ij}$ is the $j$th observed log-progesterone of the $i$th cycle, $X_{ij}$ is the day in cycles and $Z_{ij}$ is an indicator taking 1 for the conceptive woman and 0 for non-conceptive woman. $\eeta=(\alpha_0,\alpha_1,\beta_1,\cdots,\beta_K,\gamma)\trans$ are the unknown coefficients and $(t_1,\cdots,t_K)$  are the unknown locations of kink points. We consider different quantiles  $\tau=0.1, 0.25,0.5, 0.75$ and $0.9$.

We first check the existence of kink points by using the proposed test with 500 bootstrap replicates for all quantiles. The resulting P-values are all nearly 0, suggesting a highly significant kink change for the slope of the day in cycle. {We then select the number of kink points by using the Schwarz-type information criterion and identify two kink points for both methods. Table \ref{tab:em} summarizes the parameter estimation results of two proposed estimation methods at different quantiles. In general, the working independence (WI) estimators are quite close to the GEE estimators with incorporating the working within-subject correlations (WC).
In most cases, the WC estimators can achieve smaller standard errors than the WI estimators, which shows that the efficiency gain can be obtained by considering the within-subject correlations in the longitudinal data analysis.}

Based on the estimation results, we obtain the following main findings. It is obvious that two  kink points are detected for all quantiles  and  divide the day in the menstrual cycle into three stages. At the first stage, the log-progesterone values stay almost unchanged  with the increase of the day, as the estimated  $\widehat{\alpha}_1$'s are close to 0 for all quantiles. In the second stage, the  log-progesterone experiences a significant rise ($\widehat{\beta}_1+\widehat{\alpha}_1>0$ for all quantiles) after  the first estimated kink point around -0.3  to -1.3. At the location around 4.5 to 5.5, the relationship between the log-progesterone and the day in cycle  experiences a new structural change once again.
 After the second kink points, it seems that the progesterone values for the women decrease at lower quantiles $\tau=0.1,0.25$ while it goes back to be stable
for upper quantiles ($\tau=0.75,0.9$). Figure \ref{fig:2} highlights the above finding through the fitted quantile curves.
In conclusion, the progesterone level remains stable before the day of ovulation, then increases quickly in five to six days after ovulation and then changes to stable again or even drops slightly after the second kink point.

 \begin{table}
\centering
\begin{threeparttable}
\footnotesize
\small
\caption{The estimated parameters and their standard errors (in parentheses) for the  progesterone data.  P-values are computed by using the proposed score test based on 500 bootstrap replicates.  C.I. denotes the 95\% Wald-type confidence interval.}
\label{tab:em}
\begin{tabular}{clccccc}
\hline
\hline
\multirow{2}{*}{Method} & $\tau$ &  $0.1$  & $0.25$ & $0.5$ & $0.75$ & $0.9$ \\
\cmidrule(lr){2-7}
 &  P-values& 0.005 & 0.000 & 0.000 & 0.000& 0.000\\
\hline
 WI &$\widehat{K}$ & 2& 2 & 2 & 2 & 2 \\
 \cmidrule(lr){2-7}
&$\widehat{\alpha}_0$ &	 $-2.071_{(0.221)}$  &	$-1.297_{(0.136)}$ 	& $-0.760_{(0.110)}$ & $-0.212_{(0.110)}$ & $0.234_{(0.118)}$ \\
& C.I.   & [-2.505, -1.637] & [-1.565, -1.030] &  [-0.977, -0.544] & [-0.427, 0.004] &  [ 0.003,  0.465] \\
&$\widehat{\alpha}_1$ &	$-0.025_{(0.028)}$ 	&	$0.010_{(0.014)}$ 	& $0.002_{(0.011)}$ &	$0.010_{(0.087)}$ 	& $-0.014_{(0.031)}$ 	\\
& C.I.  &  [-0.080,  0.031] &   [-0.019,  0.039] &  [-0.019,  0.023] & [-0.024,  0.044] &  [-0.074,  0.047] \\
& $\widehat{\beta}_1$  & $0.480_{(0.092)}$  & $0.379_{(0.024)}$ 	& $0.381_{(0.018)}$ &	$0.438_{(0.041)}$ 	& $0.362_{(0.031)}$ 	\\
& C.I.  & [ 0.300,  0.661] & [ 0.333,  0.426] &  [ 0.345,  0.417] & [ 0.357,  0.518] &  [ 0.296,  0.427] \\
& $\widehat{\beta}_2$  & $-0.500_{(0.108)}$  & $-0.444_{(0.028)}$ 	& $-0.445_{(0.026)}$ &	$-0.441_{(0.037)}$ 	& $-0.348_{(0.038)}$ 	\\
& C.I. & [-0.712, -0.288] &  [-0.499, -0.390] &  [-0.495, -0.395] & [-0.513, -0.369] &  [-0.422, -0.274] \\
& $\widehat{\gamma}$  &  $0.331_{(0.377)}$  & $0.256_{(0.163)}$ 	& $0.205_{(0.146)}$ &	$0.124_{(0.177)}$ 	& $0.016_{(0.359)}$ 	\\
& C.I. & [-0.407,  1.069] & [-0.064,  0.575] &  [-0.082,  0.491] & [-0.223,  0.471] &  [-0.687,  0.719] \\
& $\widehat{t}_1$ &  $-0.271_{(0.541)}$  & $-0.824_{(0.203)}$ 	& $-0.925_{(0.211)}$ &	$-0.390_{(0.329)}$ 	& $-1.376_{(0.390)}$ 	\\
& C.I. & [-1.331,  0.790]   & [-1.222, -0.426] &  [-1.339, -0.510] & [-1.036,  0.255] &  [-2.141, -0.612] \\
&$\widehat{t}_2$ & $5.029_{(0.420)}$  & $5.516_{(0.312)}$ 	& $5.666_{(0.148)}$ &	$4.548_{(0.223)}$ 	& $5.319_{(0.686)}$ 	\\
&C.I. & [ 4.205,  5.853]  & [ 4.904,  6.127] &  [ 5.377,  5.956] & [ 4.110,  4.986] &  [ 3.975,  6.662] \\
\hline
WC &$\widehat{K}$ & 2& 2 & 2 & 2 & 2 \\
\cmidrule(lr){2-7}
&	$\widehat{\alpha}_0$	&	$-2.703_{(0.195)}$	&	$-1.786_{(0.130)}$	&	$-0.764_{(0.073)}$	&	$-0.081_{(0.114)}$	&	$0.817_{(0.157)}$	\\
&	C.I.	&	[-3.086, -2.321]	&	[-2.042, -1.531]	&	[-0.908, -0.620]	&	[-0.304, 0.142]	&	[ 0.510,  1.124]	\\
&	$\widehat{\alpha}_1$	&	$-0.062_{(0.028)}$	&	$-0.019_{(0.016)}$	&	$-0.005_{(0.008)}$	&	$0.003_{(0.016)}$	&	$-0.042_{(0.023)}$	\\
&	C.I.	&	[ -0.117,  -0.007]	&	[-0.050,  0.012]	&	[-0.020, 0.010]	&	[-0.027,  0.034]	&	[-0.087,  0.002]	\\
&	$\widehat{\beta}_1$	&	$0.544_{(0.096)}$	&	$0.415_{(0.038)}$	&	$0.397_{(0.012)}$	&	$0.460_{(0.038)}$	&	$0.374_{(0.025)}$	\\
&	C.I.	&	[ 0.356,  0.733]	&	[ 0.341,  0.490]	&	[ 0.373,  0.422]	&	[ 0.385,  0.535]	&	[ 0.325,  0.424]	\\
&	$\widehat{\beta}_2$	&	$-0.535_{(0.089)}$	&	$-0.477_{(0.043)}$	&	$-0.473_{(0.019)}$	&	$-0.508_{(0.033)}$	&	$-0.295_{(0.025)}$	\\
&	C.I.	&	[-0.710, -0.361]	&	[-0.562, -0.392]	&	[-0.510, -0.436]	&	[-0.572, -0.443]	&	[-0.344, -0.246]	\\
&	$\widehat{\gamma}$	&	$0.593_{(0.271)}$	&	$0.650_{0.173)}$	&	$0.279_{(0.130)}$	&	$1.114_{(0.267)}$	&	$-0.146_{(0.286)}$	\\
&	C.I.	&	[ 0.063,  1.124]	&	[ 0.311,  0.989]	&	[ 0.025, 0.534]	&	[ 0.591,  1.638]	&	[-0.706,  0.414]	\\
&	$\widehat{t}_1$	&	$-0.471_{(0.429)}$	&	$-0.774_{(0.336)}$	&	$-0.889_{(0.131)}$	&	$-0.674_{(0.193)}$	&	$-1.328_{(0.352)}$	\\
&	C.I.	&	[-1.313,  0.370]	&	[-1.432, -0.116]	&	[-1.145,  -0.632]	&	[-1.053, -0.295]	&	[-2.019, -0.638]	\\
&	$\widehat{t}_2$	&	$4.933_{(0.381)}$	&	$5.378_{(0.180)}$	&	$5.605_{(0.082)}$	&	$4.580_{(0.189)}$	&	$5.389_{(0.269)}$	\\
&	C.I.	&	[ 4.186,  5.681]	&	[ 5.025,  5.730]	&	[ 5.444,  5.766]	&	[ 4.209,  4.951]	&	[ 4.861,  5.917]	\\
\hline
\end{tabular}
\end{threeparttable}
\end{table}

\section{Conclusion}
In this article,  we develop an estimation and inference framework for the multi-kink quantile regression (MKQR) for the longitudinal data. There are also some interesting extension topics. First,
since we study the asymptotic normality properties for both estimators given the true number of kink points.  However, how to derive the asymptotic properties for estimators under mispecification and make relevant robust inference deserve to be further investigated. Second, different quantile levels may share the same kink points, so it is interesting to investigate the composite estimator for kink points to improve the efficiency. Third, how to extend the proposed method to deal with the missing data is also deserved to be further studied. The inverse probability weighting method \citep{little2019statistical}  may be helpful. Fourth, we apply the QIF approach to account for the correlation structure. One may consider some other methods to depict the within-subject dependence structures such as copula technique \citep{wang2019copula}. We will explore these issues in the future study.


\renewcommand{\baselinestretch}{1.00}
\baselineskip=14pt

\bibliographystyle{apalike}
\bibliography{mkqrl}

\end{document}